# New basis functions for wave equation

## SINA KHORASANI[*]

*School of Electrical Engineering, Sharif University of Technology, Tehran, Iran*
*\*Corresponding author: khorasani@sina.sharif.edu*

**The Differential Transfer Matrix Method is extended to the complex plane, which allows dealing with singularities at turning points. The result for real-valued systems are simplified and a pair of basis functions is found. These bases are a bit less accurate than WKB solutions but much easier to work with because of their algebraic form. Furthermore, these bases exactly satisfy the initial conditions and may go over the turning points without the divergent behavior of WKB solutions. The findings of this paper allow explicit evaluation of eigenvalues of confined modes with high precision, as demonstrated by few examples.**

***Keywords:*** *Physical optics; Quantum optics; Electromagnetic optics; Inhomogeneous optical media.*

## I. Introduction

Many physical problems in optics and quantum mechanics are modeled using the equation [1-3]

$$y''(x) + f(x)y(x) = 0 \qquad (1)$$

where $f(x) = g(x) + ih(x)$ and $y(x) = u(x) + iv(x)$ are complex analytic functions. Equation (1) may be written as

$$\frac{d}{dx}\{\mathbf{F}(x)\} = [\mathbf{K}(x)]\{\mathbf{F}(x)\} \qquad (2)$$
$$\{\mathbf{F}(x)\}^t = \{u(x)\ v(x)\ u'(x)\ v'(x)\}$$
$$[\mathbf{K}(x)] = \begin{bmatrix}[\mathbf{0}] & [\mathbf{1}] \\ [\mathbf{E}(x)] & [\mathbf{0}]\end{bmatrix}$$
$$[\mathbf{E}(x)] = -\begin{bmatrix} g(x) & -h(x) \\ h(x) & g(x) \end{bmatrix}$$

in which $[\mathbf{0}]$ and $[\mathbf{1}]$ are respectively the zero and identity matrices. Using the Differential Transfer Matrix Method (DTMM), the DTMM-like system (2) is subject to the initial conditions

$$\{\mathbf{F}(0)\}^t = \{\Re f(0)\ \Im f(0)\ \Re f'(0)\ \Im f'(0)\} \qquad (3)$$

having the *exact* solution using the DTMM [4,5]

$$\{\mathbf{F}(x)\} = \mathbb{T}\exp\left[\int_0^x [\mathbf{K}(t)]dt\right]\{\mathbf{F}(0)\} = [\mathbf{Q}_{0\to x}]\{\mathbf{F}(0)\} \qquad (4)$$

The transfer matrices $[\mathbf{Q}_{p\to q}]$ observe [4] the self-projection $[\mathbf{Q}_{p\to p}] = [\mathbf{1}]$, inversion $[\mathbf{Q}_{p\to q}] = [\mathbf{Q}_{q\to p}]^{-1}$, determinant $|\mathbf{Q}_{p\to q}| = \exp(\text{tr}\{[\mathbf{K}_{p\to q}]\})$, and decomposition $[\mathbf{Q}_{p\to r}] = [\mathbf{Q}_{q\to r}][\mathbf{Q}_{p\to q}]$ properties. One may omit the ordering operator $\mathbb{T}$ [5-7] to reach the explicit but approximate solution

$$\{\mathbf{F}(x)\} = \exp\left[\int_0^x [\mathbf{K}(t)]dt\right]\{\mathbf{F}(0)\} \cong [\mathbf{Q}_{0\to x}]\{\mathbf{F}(0)\} \qquad (5)$$

while the trace $\text{tr}\{[\mathbf{Q}_{0\to x}]\}$ [7] and the first above three properties remain intact. We now define

$$[\mathbf{M}(x)] = \int_0^x [\mathbf{K}(t)]dt = \begin{bmatrix} [\mathbf{0}] & x[\mathbf{1}] \\ [\mathbf{B}(x)] & [\mathbf{0}] \end{bmatrix} \qquad (6)$$
$$[\mathbf{B}(x)] = \int_0^x [\mathbf{E}(t)]dt = -\begin{bmatrix} G(x) & -H(x) \\ H(x) & G(x) \end{bmatrix}$$

The matrix exponentials here can be simplified as

$$[\mathbf{Q}_{0\to x}] = \cosh[\mathbf{D}(x)]\begin{bmatrix}[\mathbf{1}] & [\mathbf{0}] \\ [\mathbf{0}] & [\mathbf{1}]\end{bmatrix} + \sinh[\mathbf{D}(x)]\begin{bmatrix}[\mathbf{0}] & x[\mathbf{D}(x)]^{-1} \\ \frac{1}{x}[\mathbf{D}(x)] & [\mathbf{0}]\end{bmatrix} \qquad (7)$$

with $\sinh(\cdot)$ and $\cosh(\cdot)$ defined according to their Taylor expansions. Here, we have assumed that $[\mathbf{D}(x)]$ is a matrix root of $[\mathbf{B}(x)]$ in such a way that $x[\mathbf{B}(x)] = [\mathbf{D}(x)]^2$. If $q_{0\to x}^{ij}$ are elements of $[\mathbf{Q}_{0\to x}]$, then the solution will be

$$y(x) = [q_{0\to x}^{11}u(0) + q_{0\to x}^{12}v(0) + q_{0\to x}^{13}u'(0) + q_{0\to x}^{14}v'(0)] + i[q_{0\to x}^{21}u(0) + q_{0\to x}^{22}v(0) + q_{0\to x}^{23}u'(0) + q_{0\to x}^{24}v'(0)] \qquad (8)$$

We also define $[\mathbf{C}(x)] = \cosh[\sqrt{x}\mathbf{D}(x)] = [C_{ij}(x)]$ and $[\mathbf{S}(x)] = x[\mathbf{D}(x)]^{-1}\sinh[\mathbf{D}(x)] = [S_{ij}(x)]$ to ultimately obtain the function $y(x) = u(x) + iv(x)$ as

$$u(x) = C_{11}(x)u(0) + C_{12}(x)v(0) + S_{11}(x)u'(0) + S_{12}(x)v'(0)$$

$$v(x) = C_{21}(x)u(0) + C_{22}(x)v(0) + S_{21}(x)u'(0) + S_{22}(x)v'(0) \tag{9}$$

with the derivative $y'(x) = u'(x) + iv'(x)$

$$u'(x) = T_{11}(x)u(0) + T_{12}(x)v(0) + C_{11}(x)u'(0) + C_{12}(x)v'(0)$$
$$v'(x) = T_{21}(x)u(0) + T_{22}(x)v(0) + C_{21}(x)u'(0) + C_{22}(x)v'(0) \tag{10}$$

where

$$[\mathbf{T}(x)] = \frac{1}{x}[\mathbf{B}(x)][\mathbf{S}(x)] = [T_{ij}(x)] \tag{11}$$

## 2. Basis Functions

For real-valued equations having the form

$$u''(x) + g(x)u(x) = 0 \tag{12}$$

the DTMM [4,8] fails at zeros of $g(x)$. While the generalized DTMM [5] and Airy functions [9] are able to alleviate some problems connected to these singularities, the method discussed in this paper easily removes singular points since they are now located at zeros of $\int_0^x g(t)dt$ instead of $g(x)$. In fact, the only singular expression now corresponds to $[\mathbf{S}(x)]$, which may be shown that it could be also smoothed out, too. Hence, in real-plane with $v(0) = v'(0) = 0$, the DTMM solution is given by

$$u(x) = C(x)u(0) + S(x)u'(0) \tag{13}$$

We obtain after significant algebra [10] the new basis functions

$$C(x) = \cos\sqrt{x \int_0^x k^2(t)dt}$$
$$S(x) = x\,\text{sinc}\sqrt{x \int_0^x k^2(t)dt} \tag{14}$$

where $g(x) = k^2(x)$ and $\text{sinc}(x) = \frac{1}{x}\sin(x)$. In quantum mechanics we have $k(x) = \sqrt{\frac{2m}{\hbar^2}[E - V(x)]}$ where $m$ is mass, $\hbar$ is reduced Planck's constant, $E$ is energy, and $V(x)$ is the potential. In optical problems, we have $k(x) = c^{-2}\sqrt{\omega^2[\epsilon(x) - N]}$ where $c$ is the speed of light in vacuum, $\omega$ is the angular frequency, $\epsilon(x)$ is the relative permittivity profile of the dielectric, and $N$ is the normalized propagation constant.

It is easy to verify that if $g(x) = g(-x)$ then $C(x) = C(-x)$ and $S(x) = -S(-x)$. These functions may be shown to exactly satisfy the initial conditions $C(0) = S'(0) = 1$ and $C'(0) = S(0) = 0$.

These basis functions are not necessarily orthonormal except for the trivial case of constant wavefunction $k(x)$. They furthermore do not span a complete space, just in the same way as for a fixed $k$, $\sin(kx)$ and $\cos(kx)$ do not. But they may be always combined linearly to satisfy any arbitrary initial or boundary conditions.

In contrast, then well-known WKB bases are

$$\tilde{C}(x) = \frac{1}{\sqrt{k(x)}}\cos(\int_0^x k(t)dt)$$
$$\tilde{S}(x) = \frac{1}{\sqrt{k(x)}}\sin(\int_0^x k(t)dt) \tag{15}$$

While WKB bases (15) clearly diverge at turning points with $k(x) = 0$, the newly introduced bases (14) do not. The reason is that at the turning points where $k^2(x)$ changes sign, the relationships for these bases (14) remain algebraically continuous and differentiable. In contrast, WKB basis functions (15) are neither differentiable nor continuous at the turning points, because of the $1/\sqrt{k(x)}$ prefactor.

It is has to be pointed out, that another pair of improved basis functions using Airy functions also exists. These are actually found from asymptotic expansions near a turning point at $x = \xi$, given by [11, p. 165]

$$\bar{C}(x) = \frac{\left[\int_x^\xi k(t)dt\right]^{\frac{1}{6}}}{\sqrt{k(x)}} \text{Ai}\left\{\pm\frac{3}{2}\left[\int_x^\xi k(t)dt\right]^{\frac{2}{3}}\right\}$$

$$\bar{S}(x) = \frac{\left[\int_x^\xi k(t)dt\right]^{\frac{1}{6}}}{\sqrt{k(x)}} \text{Bi}\left\{\pm\frac{3}{2}\left[\int_x^\xi k(t)dt\right]^{\frac{2}{3}}\right\}$$

$$\tag{16}$$

where the plus or minus signs are used across the turning point respectively in the classically forbidden and allowed zones, and $\text{Ai}(\cdot)$ and $\text{Bi}(\cdot)$ are respectively the Airy's function of the first and second kind.

While relatively accurate, these bases are too complicated for practical analytical purposes and also have to be exchanged across the singularities for smooth solutions. Furthermore, it is impossible to obtain the analytical spectrum of confined systems, even as simple as harmonic oscillator using the above.

## 3. Examples

### 3.1. Periodic systems

For periodic system with $g(x) = g(x + L)$, one may adapt DTMM to calculate the Bloch waves [3,5,8], which satisfy

$$u(x; \kappa) = \exp(ix\kappa)\Theta(x; \kappa)$$
$$\Theta(x; \kappa) = \Theta(x + L; \kappa) \tag{17}$$

where $\kappa$ is the Bloch wavenumber. The DTMM solution of (17) takes the form

$$\{\mathbf{F}(x + L)\} = [\mathbf{Q}_{x \to x+L}]\{\mathbf{F}(x)\} \tag{18}$$

while the periodic boundary conditions demand

$$u(x + L) = \exp(i\kappa L)u(x)$$
$$u'(x + L) = \exp(i\kappa L)u'(x) \quad (19)$$

Finally, the simple result for the Bloch wave number $\kappa$ is

$$\exp(i\kappa L) = \text{eig}[\mathbf{Q}_{x \to x+L}] \quad (20)$$

It is not difficult to see that the extended solutions based on the functions (14) as

$$u(x;\kappa) = \exp\left(\pm i \sqrt{x \int_0^x k^2(t)dt}\right) \quad (21)$$

are actually Bloch waves. To verify this, we may write

$$\frac{u(x+L;\kappa)}{u(x;\kappa)} = \exp\left(\pm i \sqrt{(x+L-x)\int_x^{x+L} k^2(t)dt}\right)$$
$$= \exp\left(\pm i \sqrt{L \int_0^L k^2(t)dt}\right) \quad (22)$$

Comparing to (19) reveals that the Bloch wave number is actually given by the dispersion relation

$$\kappa = \frac{1}{\sqrt{L}} \sqrt{\int_0^L k^2(t)dt} \quad (23)$$

or

$$\kappa^2 = \frac{1}{L} \int_0^L k^2(t)dt \quad (24)$$

Doing the same procedure with the WKB bases leads to a different dispersion equation as

$$\kappa = \frac{1}{L} \int_0^L k(t)dt \quad (25)$$

The dispersion equation using the new basis functions (24) matches the long-wave length limit (alternatively known as homogenization) of the photonic crystals [12], while the WKB solution (25) does not. In that sense, the proposed basis functions provide higher accuracy for such class of problems.

*3.2. Confined bounded states*

We here demonstrate the power of new bases in analytical solution of a wide class of problems in optics and quantum mechanics. We suppose that an even confining potential or refractive index profile is given which is of the form

$$V(x) = U|x|^\alpha \quad (26)$$

where both $\alpha$ and $U$ are any positive real constants. The turning points are therefore located at $\xi = \pm(E/U)^{1/\alpha}$. The cases $\alpha = 2$ [1,11,13,14] and $\alpha = 1$ [14] respectively correspond to the harmonic oscillator and quarkonium potentials. Hence, the wavenumber function is defined as

$$k^2(x) = \frac{2m}{\hbar^2}[E - V(x)] \quad (27)$$

Noting the form of bases (14), the eigenstates may be found by the Wilson-Sommerfeld's quantization [14] as

$$\sqrt{2\xi \int_{-\xi}^{\xi} k^2(x)dx} = \pi\left(n + \frac{1}{2}\right) \quad (28)$$

in which $n$ is a non-negative integer denoting the state number. The $\pi/2$ phase shift on the right-hand-side of the above has to be normally added because of the reflection phase of wavefunction at the turning point.

In contrast, the WKB quantization will lead to the alternative form

$$\int_{-\xi}^{\xi} k(x)dx = \pi\left(n + \frac{1}{2}\right) \quad (29)$$

While neither (28) nor (29) are generally exact, nevertheless the WKB quantization (29) cannot be analytically integrated except for $\alpha = 1,2$. It is known that, only for the special case of $\alpha = 2$, however, the WKB quantization leads to exact result. Anyhow, plugging (26) and (27) into (28) and noting the even symmetry gives

$$4\xi \int_0^\xi \frac{2m}{\hbar^2}[E_n - Ux^\alpha]dx = \pi^2\left(n + \frac{1}{2}\right)^2 \quad (30)$$

which after integration and some simplification leads to the algebraic equation

$$\frac{\alpha U}{\alpha+1}\left(\frac{E_n}{U}\right)^{1+\frac{2}{\alpha}} = \frac{\pi^2 \hbar^2}{8m}\left(n + \frac{1}{2}\right)^2 \quad (31)$$

Hence, the $n$-th energy eigenstate will be given by

$$E_n = \left[\frac{\pi^2 \hbar^2}{8m}\left(1 + \frac{1}{\alpha}\right)U^{\frac{2}{\alpha}}\right]^{\frac{\alpha}{\alpha+2}} \left(n + \frac{1}{2}\right)^{\frac{2\alpha}{\alpha+2}} \quad (32)$$

Therefore, the ground state energy according to (32) will be

$$E_0 = \left[\frac{\pi^2 \hbar^2}{8\sqrt{2}m}\left(1 + \frac{1}{\alpha}\right)U^{\frac{2}{\alpha}}\right]^{\frac{\alpha}{\alpha+2}} \quad (33)$$

It is easy to verify that the expression for spectrum (32) indeed agrees remarkably well with the known behavior of spectrum for quarkonium $E_n \sim \left(n + \frac{1}{2}\right)^{2/3}$ and harmonic oscillator $E_n \sim n + \frac{1}{2}$.

For the case of quarkonium with $\alpha = 1$, we get from (32)

$$E_n = \sqrt[3]{\frac{\pi^2\hbar^2 U^2}{4m}}\left(n+\frac{1}{2}\right)^{\frac{2}{3}} \qquad (34)$$

while the WKB solution [14] is

$$E_n = \sqrt[3]{\frac{9\pi^2\hbar^2 U^2}{32m}}\left(n+\frac{1}{2}\right)^{\frac{2}{3}} \qquad (35)$$

These relations differ within a factor of $2/\sqrt[3]{9}$, corresponding to an error of only 3.8%.

Also, for the case of harmonic oscillator with $\alpha = 2$, and by taking $\frac{1}{2}m\Omega^2 = U$, the spectrum according to the proposed bases is

$$E_n = \sqrt{\frac{3\pi^2}{32}}\,\hbar\Omega\left(n+\frac{1}{2}\right) \qquad (36)$$

which differs with the exact spectrum $E_n = \hbar\Omega\left(n+\frac{1}{2}\right)$ within a factor of $\sqrt{3\pi^2/32}$ corresponding to an error of again 3.8%.

It is also instructive to check out the limiting case of $\alpha \to \infty$ in (32). We first note that the wavefunction should abruptly terminate at the turning points, because of the existence of the infinite potential wall at $x = \pm 1$. Hence, the $\pi/2$ phase shift on the right-hand-side of (28) would be unnecessary. Now, taking the limit here yields

$$E_n = \frac{\pi^2\hbar^2}{8m}n^2 \qquad (37)$$

which is exactly the well-known spectrum of a confined massive particle in infinite potential well.

### 3.3. Singular potentials

As another example, we may also verify the existence of bounded solutions for the singular even confining potential of the form

$$V(x) = -U|x|^{-\beta} \qquad (38)$$

where $\beta < 1$ is any positive real constant less than 1. The turning points are now located at $\xi = \pm(-U/E)^{1/\beta}$. Using (28) results in

$$\xi \int_0^\xi (E_n + Ux^{-\beta})dx = \frac{\pi^2\hbar^2}{8m}\left(n+\frac{1}{2}\right)^2 \qquad (39)$$

offering the solution for the energy spectrum as

$$E_n = -\left[\left(\frac{1}{\beta}-1\right)\frac{\pi^2\hbar^2}{8U^{2/\beta}m}\right]^{\frac{\beta}{\beta-2}}\left(n+\frac{1}{2}\right)^{\frac{2\beta}{\beta-2}} \qquad (40)$$

Since $\beta - 2$ is negative, then it would be better to rewrite the above as

$$E_n = -\left[\frac{8\beta U^{2/\beta}m}{\pi^2\hbar^2(1-\beta)}\right]^{\frac{\beta}{2-\beta}}\frac{1}{\left(n+\frac{1}{2}\right)^{\frac{2\beta}{2-\beta}}} \qquad (41)$$

showing that $E_n \sim 1/\left(n+\frac{1}{2}\right)^{\frac{2\beta}{2-\beta}}$. Therefore, the ground state corresponding to the potential (38) is now given by

$$E_0 = -\left[\frac{32\beta U^{2/\beta}m}{\pi^2\hbar^2(1-\beta)}\right]^{\frac{\beta}{2-\beta}} \qquad (42)$$

which is divergent when $\beta \to 1$. This justifies the existence of a ground state bounded from below only when $0 < \beta < 1$.

#### 3.3.1. Numerical Example

For the case of $\beta = \frac{1}{2}$ in the normalized atomic units where $\hbar^2/2m \to 1$ and $U \to 1$ we obtain from (40)

$$E_n = -\left[\frac{\pi}{2}\left(n+\frac{1}{2}\right)\right]^{-\frac{2}{3}} \qquad (43)$$

Hence, the energy of the ground and first excited state are respectively $E_0 = -\sqrt[3]{16/\pi^2} \sim -1.17474$ and $E_1 = -\sqrt[3]{16/9\pi^2} \sim -0.56475$. The exact values by numerical computation are $E_0 = -1.6534$ and $E_1 = -0.43804$, thus the error of estimation (43) for the first two states is about 28.9%. Figure 1 illustrates the numerically calculated wavefunctions and the eigenstates inside the quantum well.

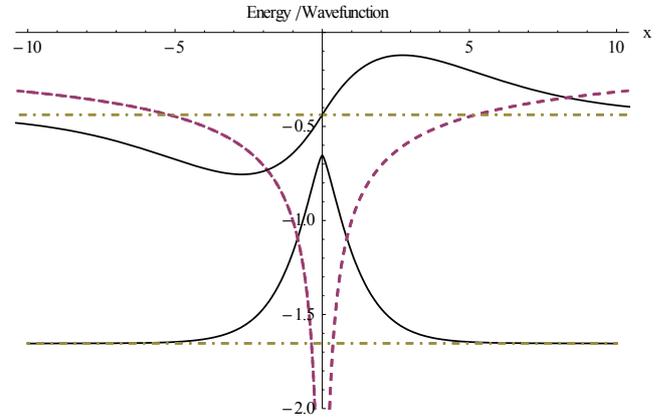

Fig. 1. Illustration of numerically calculated wavefunctions (solid black) inside the singular quantum well (dashed) with $\beta = \frac{1}{2}$. The ground and first excited eigenstates are leveled with horizontal lines (dot-dashed).

Figure 2 compares the numerically exact solution of the unnormalized wavefunctions. Here, the ground and first excited states are shown versus the solution obtained by the our proposed bases (14), improved WKB bases (15), as well as simple WKB basis $\cos\left(\int_0^x k(t)dt\right)$. The superior accuracy of the solution by (14) is clearly visible from this plot. The improved WKB bases are highly erroneous for bounded modes as can be seen here, while simple WKB solutions are still less accurate than our proposed basis functions (14).

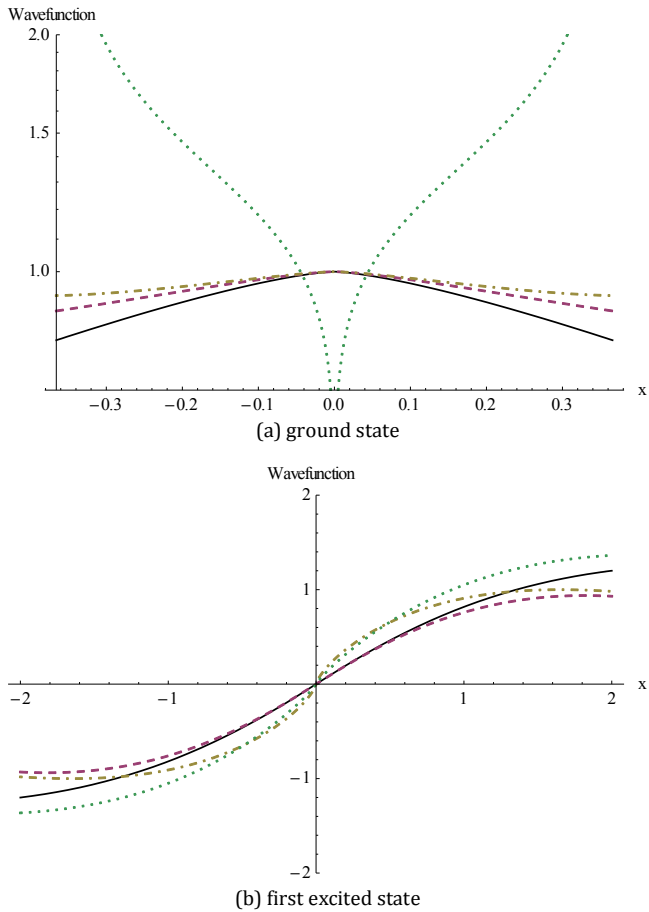

Fig. 2. Comparison of unnormalized ground state wavefunctions for $\beta = \frac{1}{2}$ within a subdomain of the range spanned by turning points $[-\xi, +\xi]$: numerically exact (solid black); newly proposed basis functions (dashed); simple WKB (dot dashed); improved WKB (dotted). All solutions diverge at infinity beyond the turning points, but the improved WKB bases diverge at the turning points.

## 4. Conclusions

In summary, we presented a new pair of bases for solving the wave equation. Through several explicit examples, we have established the analytical power of proposed basis functions and their advantage over WKB basis solutions. Our new basis functions are not divergent close to singularities at turning point, and they maintain remarkable accuracy compared to the analytical solutions. The basis functions also satisfy the initial conditions exactly.

**References**

1. A. Yariv, An Introduction to Theory and Applications of Quantum Mechanics, Dover, 2013.
2. S. Khorasani, and K. Mehrany, "Differential Transfer Matrix Method for Solution of One-dimensional Linear Non-homogeneous Optical Structures," J. Opt. Soc. Am. B, vol. 20, no. 1, pp. 91-96 (2003).
3. S. Khorasani, and A. Adibi, "New Analytical Approach for Computation of Band Structure in One-dimensional Periodic Media," Optics Communications, vol. 216, no. 4-6, pp. 439-451 (2003).
4. S. Khorasani, and A. Adibi, "Analytical Solution of Linear Ordinary Differential Equations by Differential Transfer Matrix Method," Electron. J. Diff. Eq., vol. 2003(79), pp. 1-18 (2003).
5. S. Khorasani, "Differential Transfer Matrix Solution of Generalized Eigenvalue Problems," Proc. Dynamic Systems & Appl., vol. 6, pp. 213-222 (2012).
6. R. M. Wilcox, "Exponential Operators and Parameter Differentiation in Quantum Physics," J. Math. Phys., vol. 8, pp. 962-982 (1967).
7. S. Khorasani, "Reply to Comment on 'Analytical Solution of Nonhomogeneous Anisotropic Wave Equation Based on Differential Transfer Matrices'," J. Opt. A: Pure Appl. Opt., vol. 5, pp. 434-435 (2003).
8. S. Khorasani, and A. Adibi, "New Analytical Approach for Computation of Band Structure in One-dimensional Periodic Media," Opt. Comm., vol. 216, pp. 439-451 (2003).
9. N. Zariean, P. Sarrafi, K. Mehrany, and B. Rashidian, "Differential-Transfer-Matrix Based on Airy's Functions in Analysis of Planar Optical Structures With Arbitrary Index Profiles," IEEE J. Quant. Electron., vol. 44, pp. 324-330 (2008).
10. S. Khorasani and F. Karimi, "Basis functions for solution of non-homogeneous wave equation," Proc. SPIE, vol. 8619, 86192B (2013).
11. W. Schleich, Quantum Optics in Phase Space, While-VCH, Berlin (2001).
12. S. T. chui, Z. F. Lin, "Long-wavelength behavior of two-dimensional photonic crystals," Phys. Rev. E, vol. 78, 065601(R) (2008).
13. S. Khorasani, Applied Quantum Mechanics (in Persian), Delarang, Tehran (2010).
14. J. J. Sakurai, Modern Quantum Mechanics, rev. ed., Addison-Wesley, 1993.